\documentclass[twocolumn,superscriptaddress,showpacs,preprintnumbers,floatfix,aps]{revtex4-1}

\usepackage[pdftex]{color,graphicx}
\usepackage{latexsym,amsmath,amssymb,gensymb}
\usepackage{hyperref}

\begin{document}

\title{Disentangling time-focusing from beam divergence: a novel approach for high-flux thermal neutron spectroscopy at continuous and long-pulse sources}

\author{M. Zanatta}
\email{marco.zanatta@univr.it}
\affiliation{Dipartimento di Informatica, Universit\`a di Verona, Strada le Grazie 15, 37134 Verona, Italy}

\author{K.H. Andersen}
\affiliation{European Spallation Source ERIC, P.O. 176, 221 00 Lund, Sweden}

\author{P.P. Deen}
\affiliation{European Spallation Source ERIC, P.O. 176, 221 00 Lund, Sweden}

\author{A. Orecchini}
\affiliation{Dipartimento di Fisica e Geologia, Universit\`a di Perugia, Via A. Pascoli, 06123 Perugia, Italy}
\affiliation{IOM-CNR, c/o Dipartimento di Fisica e Geologia, Universit\`a di Perugia, Via A. Pascoli, I-06123 Perugia, Italy}

\author{A. Paciaroni}
\affiliation{Dipartimento di Fisica e Geologia, Universit\`a di Perugia, Via A. Pascoli, 06123 Perugia, Italy}

\author{C. Petrillo}
\affiliation{Dipartimento di Fisica e Geologia, Universit\`a di Perugia, Via A. Pascoli, 06123 Perugia, Italy}
\affiliation{INFN - Sezione di Perugia, Via A. Pascoli, 06123 Perugia, Italy}

\author{F. Sacchetti}
\affiliation{Dipartimento di Fisica e Geologia, Universit\`a di Perugia, Via A. Pascoli, 06123 Perugia, Italy}
\affiliation{IOM-CNR, c/o Dipartimento di Fisica e Geologia, Universit\`a di Perugia, Via A. Pascoli, 06123 Perugia, Italy}

\date{\today}

\begin{abstract}
We present the concept of a novel time-focusing technique for neutron spectrometers, which allows to disentangle time-focusing from beam divergence. The core of this approach is a double rotating-crystal monochromator that can be used to extract a larger wavelength band from a white beam, thus providing a higher flux at the sample compared to standard time-of-flight instruments, yet preserving energy resolution and beam collimation. The performances of a spectrometer based on this approach are quantitatively discussed in terms of possible incident wavelengths, flux at the sample and $(Q,E)$-resolution. Analytical estimates suggest flux gains of about one order of magnitude at comparable resolutions in comparison to conventional time-of-flight spectrometers. Moreover, the double monochromator configuration natively shifts the sample away from the source line-of-sight, thus significantly improving the signal-to-noise ratio. The latter, in combination with a system that does not increase the beam divergence, brings the further advantage of a cleaner access to the low-$Q$ region, which is recognized to be of fundamental interest for magnetism and for disordered materials, from glasses to biological systems.
\end{abstract}


\maketitle

\section{Introduction}

Direct geometry spectrometers for thermal and cold neutron beams are key instruments for inelastic neutron scattering (INS) experiments aimed at probing the atomic dynamics over broad intervals of exchanged momentum $Q$ and energy transfer $E$. The scientific case requiring the development of these instruments is very wide and ranges from magnetism and strongly correlated electron systems to disordered systems, soft matter and biophysics.
 
The major limitation of the INS technique, when compared with other experimental probes (e.g. synchrotron light), is given by the combination of the small inelastic neutron cross section and the intrinsically low flux available at the neutron sources. Hence, one of the main goals for an INS instrument is to maximize the flux at the sample position. Considering that the efficiency of thermal neutron sources has not shown significant increases over the last forty years, improvements to the INS technique have been brought about by a careful exploitation of the neutron beams through novel neutron optical devices, high-performance detectors and new concepts in instrument design. This paradigm holds, as well, for the forthcoming European Spallation Source (ESS, Lund, Sweden)\cite{LindroosNIMB2011,VettierNIMA2009}. The long neutron pulses of the ESS require specific optimization of the instruments in order to make full advantage of the promised increases in peak flux compared to steady-state sources and in time-average flux compared to short-pulse sources. 

The most recently developed direct-geometry spectrometers exploit coupling of a chopper cascade with a large-area position sensitive detector (PSD). This is the case of IN5 at the Institut Laue Langevin (ILL, Grenoble, France) \cite{OllivierSFN2010}, LET at ISIS (Didcot, UK) \cite{BewleyNIMA2002}, ARCS, SEQUOIA and CNCS at the Spallation Neutron Source  (SNS, Oak Ridge, USA) \cite{EhlersRSI2011,StoneRSI2014}, 4SEASON, AMATERAS and HRC (J-PARC, Japan) \cite{ KajimotoJPSJ2010, NakajimaJPSJ2011, ItohJPSJ2013}, and, in the future, T-REX\cite{VoigtNIMA2014}, CSPEC and VOR\cite{DeenJPCS2015}, at the ESS. Rotating disc choppers slice the white beam producing short monochromatic pulses that impinge on the sample and, once scattered, are analysed via the neutron time-of-flight (ToF). This technique offers great versatility, as to the selection of the incident energy and energy resolution, and benefits directly from the high peak brightness available at pulsed sources. 

For long-pulse and continuous sources, the performances can be enhanced over the thermal and cold-neutron ranges, using hybrid spectrometers exploiting the time-focusing technique. This approach consists in selecting a broader portion of the white beam in such a way that neutrons of different velocities reach the detector at the same time. Compared with standard chopper instruments, this configuration uses a longer extraction time, which provides a higher flux at the sample. However, the typical implementation of the time-focusing technique exploits an increased beam divergence to broaden the band of accepted wavelengths, therefore the flux gain is counterbalanced by a poorer $Q$-resolution. Examples of instruments employing time-focusing are IN4C\cite{CicognaniPB2000} and IN6\cite{Blanc1983}, both at ILL, and FOCUS at the Paul Scherrer Institut (PSI, Villigen, Switzerland) \cite{JuranyiCP2003}.

Here, we present a novel and alternative approach to time-focusing which is not based on a larger beam divergence. The key element of this implementation is a double rotating-crystal monochromator (DRCM) \cite{ZanattaJPCS2016} that can be used to extract a wider wavelength band from a white beam, while preserving the original beam collimation. Incident wavelength and resolution can be easily tuned by changing the rotation speed and the relative position of the DRCM elements. Consequently, a spectrometer exploiting such an approach can provide a higher flux by accepting a wider wavelength range, while achieving good resolution due to time-focusing. It does this retaining good versatility and without increasing the divergence of the beam. In addition, the DRCM removes the monochromatic beam well away from the primary white beam, which avoids the direct view of the neutron source from the sample position with a consequently significant background reduction.

\section{Basic concept}
\label{Section:Basics}
\begin{figure}[htb]
	\centering
		\includegraphics[width=0.48\textwidth]{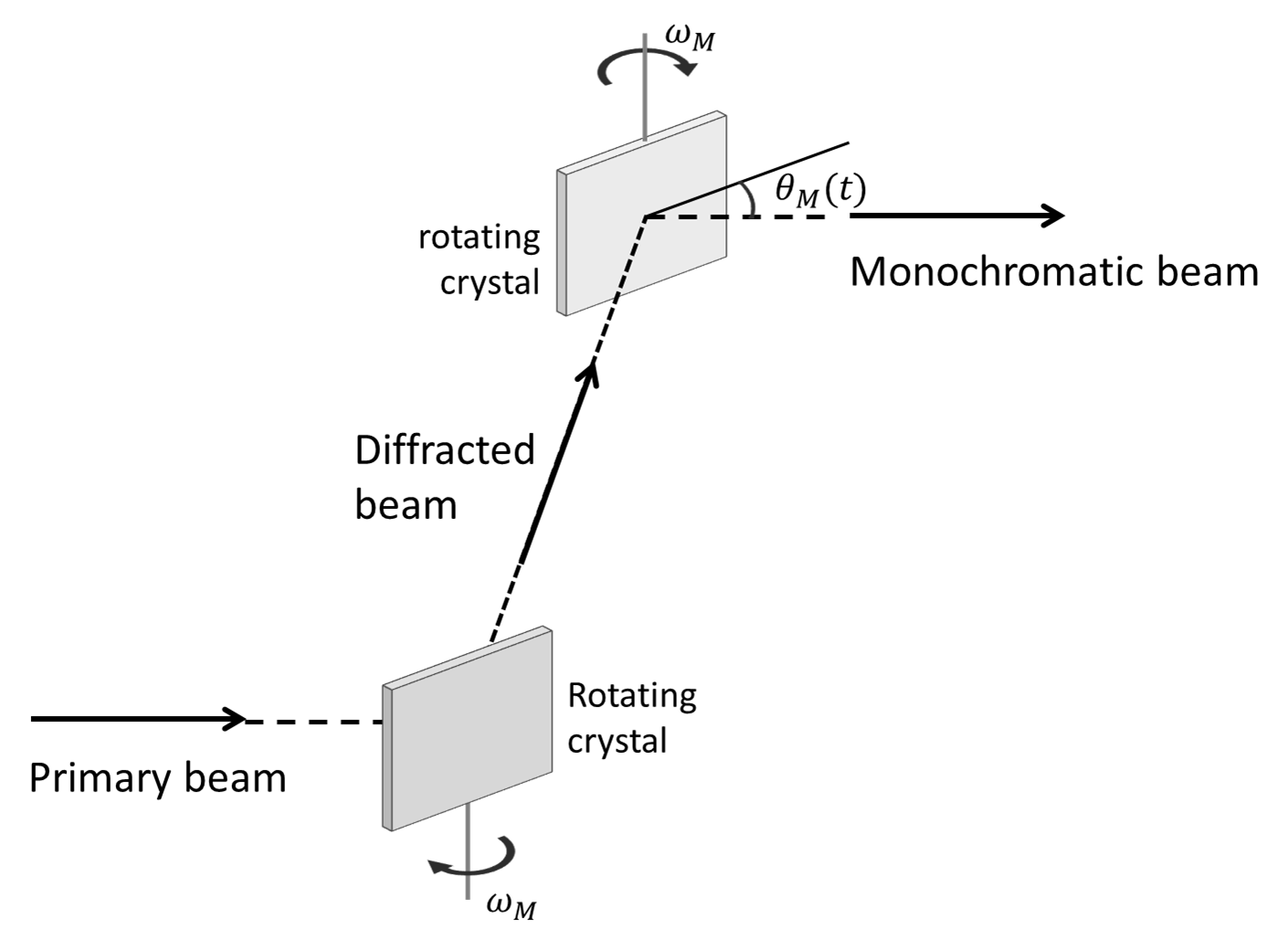}	
		\caption{Schematic view of a double rotating-crystal monochromator.}
	\label{fig1}
\end{figure}

Fig. \ref{fig1} shows a schematic view of a DRCM. The first crystal is fully bathed by the white primary beam and rotates at a given frequency $\nu_M$, with a continuous sweep of the Bragg angle $\theta_M$. The second crystal rotates in the same direction with the same frequency $\nu_M$. It is located and phased so as to properly collect the neutrons that are diffracted by the first crystal in a small range of Bragg angles and to merge them into a single parallel beam. Such a quasi-monochromatic beam is sent to the sample and finally scattered towards the detector. A proper choice of the rotation frequency allows the slowest and the fastest neutrons to catch up together at the detector position, thus meeting the time-focusing condition at a chosen energy transfer. This strategy is common to all time-focusing instruments but the use of the DRCM avoids the increase of the beam divergence as it occurs in standard approaches like, for instance, IN6 and IN4C. Indeed, in non-dispersive configuration \cite{WillisAC1960}, a double crystal monochromator transfers the divergence of the primary beam in the scattering plane to the diffracted beam with ideally no changes \cite{HohlweinJAC1988}. This is well described in Ref. \cite{RisteNIMA1969} where a specific X-ray application, with a tightly collimated primary beam, is reported. This approach results in an increased phase space volume that is proportional to the increase of the wavelength band.

\subsection{Time-focusing conditions}
For a quantitative evaluation, let us imagine to extract a rectangular pulse of duration $T$ out of the primary white beam, with $T$ of the order of hundreds of $\mu$s. This pulse impinges on the first monochromator which rotates at an angular frequency $\omega_M=2\pi \nu_M$. Setting the time origin at the middle of the collection time $T$, the Bragg angle of the rotating monochromator continuously sweeps from $\theta_M(-T/2)$ to $\theta_M(+T/2)$ in the time interval $-T/2\leq t \leq +T/2$, with a mean Bragg angle  $\theta_0=\theta_M(0)$. The time-focusing condition requires slow neutrons to be diffracted before fast ones, i.e. that the wavelength $\lambda (t)$ is a decreasing function of $t$. Consequently, the direction of rotation must satisfy the condition $\theta_M(-T/2)>\theta_M(+T/2)$. 

With $L_{MS}$ the flight path from the first-monochromator-to-sample and $L_{SD}$ the path from the sample-to-detector, the time-focusing condition imposes that
\begin{eqnarray}
\frac{L_{MS}}{v_1(-T/2)} &+& \frac{L_{SD}}{v_2(-T/2)} - T/2 \nonumber\\
&=& \frac{L_{MS}}{v_1(+T/2)} + \frac{L_{SD}}{v_2(+T/2)} + T/2
	\label{Eq:TF1}
\end{eqnarray}
\noindent
where the subscript 1 [2] refers to neutrons before [after] the sample, $v_1(t)$ [$v_2(t)$] is the velocity of neutrons with wavelength $\lambda_1(t) = h / m v_1(t)$ [$\lambda_2(t) = h / m v_2(t)$], reflected by the first monochromator at time $t$. Here $m$ is the neutron mass. Eq. \ref{Eq:TF1} is exact only when $T$ is not too large, so that the Bragg angle variation is small enough for the trigonometric functions development to be valid at the first order. We can thus define $\Delta v_i = v_i(+T/2) - v_i(-T/2) \ge 0$ and assume that $\Delta v_i \ll v_i(0)$. In this limit Eq. \ref{Eq:TF1} becomes:  
\begin{eqnarray}
\frac{L_{MS}}{v_1(0)} \frac{\Delta v_1}{v_1(0)} + \frac{L_{SD}}{v_2(0)} \frac{\Delta v_2}{v_2(0)} \simeq T.
	\label{Eq:TF2}
\end{eqnarray}

For purely elastic scattering, neutron velocities remain unchanged along the monochromator-detector path $L_{MD}=L_{MS}+L_{SD}$. Consequently, $v_1(t) = v_2(t)$ and Eq. \ref{Eq:TF2} can be written as
\begin{eqnarray}
\frac{L_{MD}}{v_1(0)}\frac{\Delta v_1}{v_1(0)} &=& - \frac{L_{MD}}{v_1(0)} \frac{\Delta \lambda}{\lambda(0)} \nonumber\\
&=& - \frac{L_{MD}}{v_1(0)} \frac{\Delta \theta_M}{\tan \theta_M(0)} = T,
	\label{Eq:TF3}
\end{eqnarray}
\noindent
where $\Delta \lambda = \lambda(+T/2) - \lambda(-T/2) \le 0$ and $\Delta \theta_M = \theta_M(+T/2) - \theta_M(-T/2) \le 0$. Writing the Bragg angle as $\theta_M(t) = \theta_M(0) + \omega_M \, t$, we obtain:
\begin{eqnarray}
\frac{L_{MD}}{v_1(0)} \, \frac{\omega_M}{\tan \theta_M(0)} = 1.
	\label{Eq:TF4}
\end{eqnarray}
Equation \ref{Eq:TF4} shows that, for a chosen neutron wavelength $\lambda$, which fixes $v_0$ and $\theta_0$, time-focusing can be achieved by adjusting the monochromator-to-detector distance $L_{MD}$ and the monochromator angular speed $\omega_M$. Considering a fixed Bragg angle $\theta_0$ and multiplying the angular speed $\omega_M$ by an integer factor $n$, Eq. \ref{Eq:TF4} is satisfied for $nv_0$, i.e. for $\lambda_0/n$. This corresponds to higher-order reflections at the same Bragg angle. Therefore, when the time-focusing condition is achieved for a given Bragg reflection, it is also achieved for all the other-order reflections, which provides a larger choice of neutron wavelengths despite the practical constraints on $L_{MD}$ and $\theta_M$. Moreover, when the  flight path between the two rotating crystals is long enough, e.g. about 1~m, their relative phase allows to select only a specific wavelength, rejecting all the other reflections.

\subsection{Energy and momentum resolution}
\label{Section:Resolution}

In general, the instrument resolution is described by a $4\times4$ matrix, e.g. Ref. \cite{VioliniNIMA2014}. However, the most important components of the resolution matrix usually considered to capture the performance of an instrument are those related to the energy transfer $E$ and the in-plane momentum transfer only. Consequently, to evaluate the potential of the proposed approach, we performed a simplified analytic calculation of the energy and momentum resolution function as a function of $E$ and scattering angle $\theta_s$. We assume a beam with an axial symmetry around the neutron transport axis, i.e. same size and divergence along all the radial directions, and a static configuration for the double monochromator, as defined at $t=0$. The latter assumption is justified if the rotation frequencies of the DRCM are such that the Doppler effect can be neglected as in the present case, see Sec. \ref{Section:Instrument}. 

When perfect time-focusing is achieved, under the assumption of a Gaussian distribution of the monochromator angles, and neglecting the effect of sample size, the energy resolution $W_{TF}$ can be written as:
\begin{eqnarray}
W_{TF} = \sqrt{(\Delta E_1 - \Delta E_2)^2 + \Delta E_D^2}.
	\label{eq:res1}
\end{eqnarray}
where the energy-dependent terms account for uncertainties in the exchanged-energy $E$ due to crystal monochromator, time-of-flight determination, and finite detector size. Equation \ref{eq:res1} is composed by two parts. The first one is $\Delta E = \Delta E_1 -\Delta E_2$, where $\Delta E_1$ and $\Delta E_2$ are the energy spread of the incident and scattered beam, respectively. The term $\Delta E$ is the direct sum of two statistically not-independent contributions, as both $E_1$ and $E_2$ depend on the wavelength of the incoming neutrons. Indeed, the neutron time of flight for sample-to-detector is obtained by subtracting the neutron monochromator-to-sample ToF from the total ToF. Consequently, the statistical fluctuations on the incoming wavelength affect the time of flight for covering both the paths $L_{MS}$ and $L_{SD}$. The second term $\Delta E_D$ derives from the time uncertainty at the detector, and it is statistically independent from the previous one. This contribution largely exceeds the contributions from the other path uncertainties because all neutron paths are almost parallel so that they have the same length at the first order in the beam divergence.

Assuming a Gaussian mosaic distribution, the energy spread $\Delta E_1$ can be easily derived from the behavior of the double monochromator. Indeed, if an ideally collimated and monchromatic beam impinges a double crystal monchromator, the intensity after the device depends on the angular acceptance of the double crystal monochromator. The transmission of the device is proportional to the coupled probability of the two crystals being close to the reflection position, that is $\left[\exp(-\log 2 \, \Delta\theta^2 /\eta^2)\right]^2$, $\Delta \theta$ being the angular distance from the Bragg position of each crystal. In other words, the two crystal device behaves like a single monochromator having a FWHM equal to $\eta/\sqrt{2}$. We can thus write:
\begin{eqnarray}
\Delta E_1 = \frac{\partial E_1}{\partial \lambda_1} \Delta\lambda_1 \frac{\eta_M}{\sqrt{2} \tan \theta_M}
\end{eqnarray}
and
\begin{eqnarray}
\Delta E_2 = \frac{\partial E_2}{\partial \lambda_1} \Delta\lambda_1 \frac{\eta_M}{\sqrt{2} \tan \theta_M}.
\end{eqnarray}

If the energy transfer is small in comparison with the incoming neutron energy, i.e. $\vert E \vert \ll E_1$, we can write:
\begin{eqnarray}
\Delta E_1 - \Delta E_2 = &-&\sqrt{2}\frac{\eta_M}{\tan \theta_M}\nonumber\\ 
&\times&\left[ E_1 \left( \frac{L_{MS}}{L_{SD}} + 1 \right) - \frac{3}{2} E \frac{L_{MS}}{L_{SD}} \right].
\label{eq:tfres}
\end{eqnarray}

These terms are dominant and other contributions, such as sample and monochromator thickness effects, are neglected. Assuming a detector thickness $l_D$ the term $\Delta E_D$ becomes
\begin{eqnarray}
\Delta E_D = &2& (E_1-E) \nonumber \\
 &\times&\sqrt{\left(\frac{l_D}{L_{SD}}\right)^2 + \frac{E_1-E}{m} \left(\frac{\eta_M}{\omega_M L_{SD}}\right)^2}.
\label{eq:detres} 
\end{eqnarray}

The above equations are valid in the limit of small energy transfer when the perfect time-focusing condition is achieved. In particular, Eqs. \ref{eq:tfres} and \ref{eq:detres} show that the monochromator-to-sample distance $L_{MS}$ should be reduced as much as possible while $L_{SD}$ should be made as long as possible. This minimizes the effect of the wavelength spread due to the mosaic $\eta_M$ as well as the detector term $\Delta E_D$.

Finally, we can provide an estimate of the energy resolution moving away from perfect time-focusing. This produces a non-random time spread term. However, considering that the neutron arrival time is random and this time spread term is not correlated to the other random contributions to $W_{TF}$, the total energy resolution $W_E$ can be written as:
\begin{eqnarray}
W_E = \sqrt{W_{TF}^2 + 4 \left[ \left(1 - \frac{E}{E_1} \right) E \frac{\omega_M T}{\tan \theta_M} \right]^2}.
\end{eqnarray}
where $T$ is the collection time of a DRCM.

In general, the determination of the $Q$-resolution implies a full calculation that includes the energy distribution and its correlation to the angular distribution. However, the DRCM provides a good disentanglement between energy and angular distributions. Moreover, in an instrument based on a crystal monochromator, the divergence of the incoming beam is usually rather tight, comparable to the crystal mosaic, with similar horizontal and vertical values at the sample position. Therefore, for the order-of-magnitude estimate, we can assume that, given a scattering angle $2\theta_s$ either horizontal or vertical, the momentum transfer in the quasi-elastic configuration is $Q = 4 \pi \sin \theta_s / \lambda_1$. Having only two independent variables $\lambda_1$ and $\theta_s$, the momentum spread is given by:
\begin{eqnarray}
W_Q = \sqrt{\left(\frac{\partial Q}{\partial \lambda_1} \Delta \lambda_1 \right)^2 + \left(\frac{\partial Q}{\partial \theta_s} \Delta \theta_s \right)^2}
\label{eq:qres1}
\end{eqnarray}
\noindent
where $\Delta \lambda_1$ and $\Delta \theta_s$ are the rms spreads of incoming wavelength and scattering angle. In the case of a small contribution from $\Delta \lambda_1$, the $Q$ resolution is reduced to:
\begin{eqnarray}
W_Q = 4 \, \pi \, {\cos \theta_s \Delta \theta_s \over \lambda_1(0)}
\label{eq:qres2}
\end{eqnarray}
\noindent
where $\theta_s$ is half of the scattering angle at the sample and $\Delta \theta_s$ is its angular spread. For a finite energy transfer, the $Q$-resolution is only slightly modified from the elastic limit.

\section{Instrument layout}
\label{Section:Instrument}
\begin{figure}[htb]
	\centering
		\includegraphics[width=0.48\textwidth]{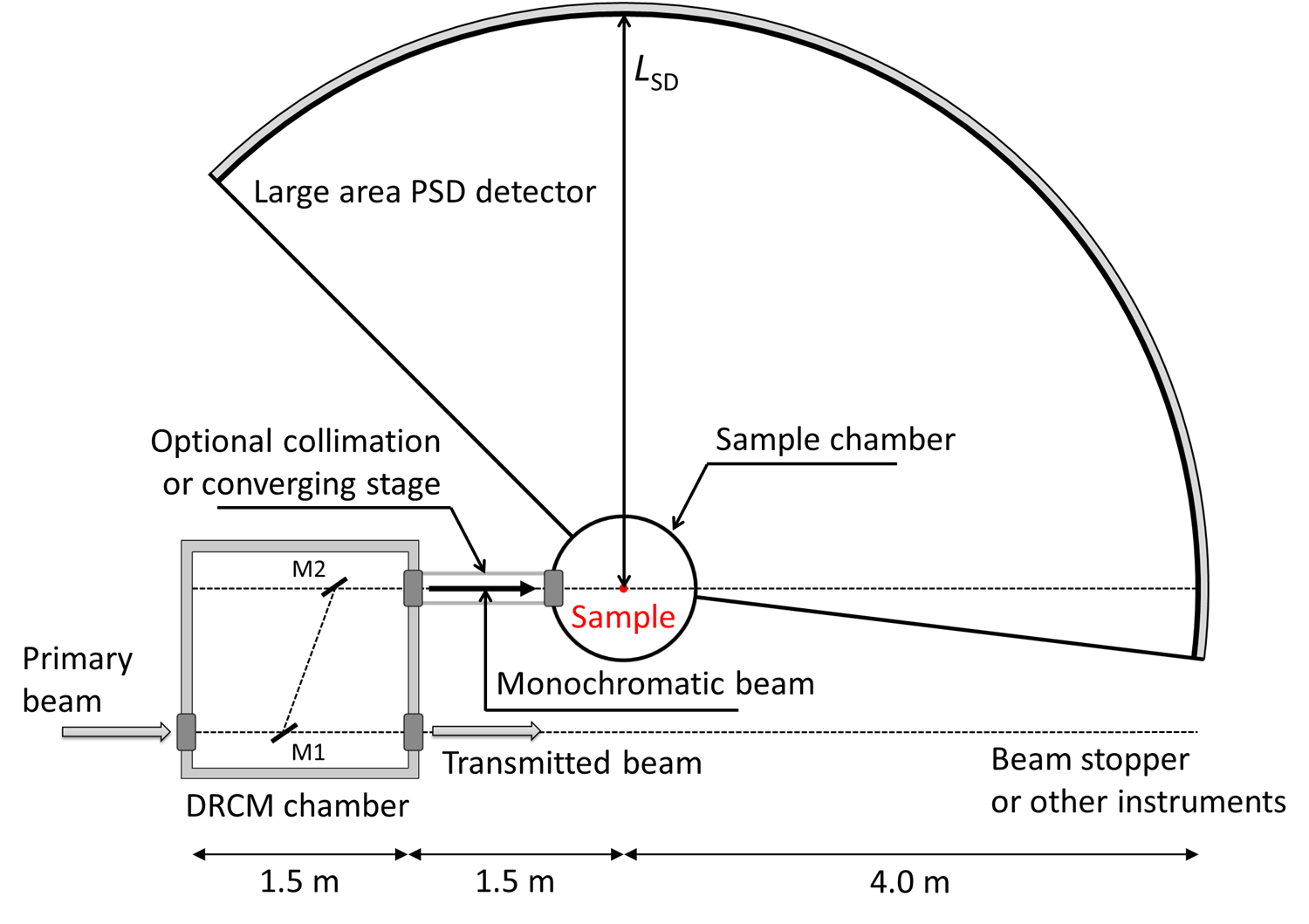}	
		\caption{Schematic view of the possible layout of an instrument based on a DRCM.}
	\label{fig2}
\end{figure}

Starting from the above considerations, we can now outline the main characteristics of an instrument based on the proposed time-focusing approach. A possible layout is shown in Fig. \ref{fig2}.

The core component is the double rotating-crystal monochromator. In order to maximize the flux at the sample, this device should cover the largest possible area, ideally of the order of $20\times20$ cm$^2$, and rotate at frequencies up to about 100 Hz (6000 rpm). 
However, the rotation of a large-surface device implies considerable mechanical difficulties that can be minimized by partitioning the large-area monochromator into a suitable number of smaller rotating elements, each containing a subset of small crystals conveniently held and aligned by a proper mechanical support. Consequently, we propose a 2cm-diameter cylindrical crystal holder composed by two symmetric halves, where slab-shaped crystals lay on the cylinder axis. A suitable number of such cylindrical elements can be easily disposed next to each other with parallel vertical axes, to cover the desired total monochromator area. The small radius of the cylinders significantly lowers the peripheral speed of the crystal edges with respect to a wholly rotating crystal, thus making rotational forces and Doppler effects on the speed of the diffracted neutrons negligible. Indeed, considering a highest rotation frequency of about 100~Hz, the peripheral speed of the crystals is about 7~m/s, compared with the lowest neutron velocity of about 1000~m/s ($\sim4$~\AA, $\sim5$~meV).

As a first possible choice for the crystals, we considered Highly Oriented Pyrolytic Graphite (HOPG) that provides a high peak reflectivity even at relatively short wavelengths ($R_0\simeq0.7$ at about 1.3~\AA, see Ref. \cite{RisteNIMA1969}), with a clean reflected beam and no spurious components due to multiple reflections. In particular, a high reflectivity is crucial to minimize intensity losses due to the double monochromator configuration. A detailed description of the construction and test of a DRCM based on these observations is reported in Ref. \cite{ZanattaJPCS2016}.

At the level of the present study, we do not wish to specialize the instrument to the characteristics of a specific neutron source. Consequently, we do not analyze in detail the primary neutron transport system, and we limit our description to the last part of the beamline, from the DRCM to the detector. In any case, in order to fully exploit its performances, the instrument should be installed downstream a guide with a large cross section. Moreover, the coupling to a crystal monochromator requires a limited beam divergence, thus for thermal neutrons with wavelengths down to 1~\AA, a supermirror guide with $m = 4$ is a good compromise.
The characteristics of the primary spectrometer are defined by the casemate containing the DRCM. In fact, the chamber has to be long enough to enable an appropriate variation of the Bragg angle $\theta_M$ and the time-focusing distance, by moving the two elements of the DRCM. This ensures great versatility to the instrument, allowing to change both wavelength and resolution. Given the high transparency of HOPG crystals off Bragg scattering, multiplexing can be implemented by adding further DRCM pairs in the chamber. 

The monochromator chamber is followed by the last portion of the beam line which brings neutrons to the sample. Due to the intrinsic geometry of the double monochromators, the monochromatic (or multichromatic, in case of multiple DRCM pairs) beam is moved away from the primary beam, so that the sample is never in direct sight of the neutron source. This is expected to result in a significant background drop and a consequent signal-to-noise improvement.

The last section of the beam line between the monochromator casemate and the sample has to include the options for a low-divergence collimation or a converging guide. This might be implemented by equipping the beam line either with mechanically interchangeable inserts or with adaptive optics. The converging-guide option is extremely useful in the case of very small samples, as it allows to concentrate the flux on the small sample volume.

The whole flight path of the primary spectrometer should be kept under vacuum. Single-crystal silicon windows, which present negligible beam attenuation off Bragg scattering, can be used to effectively separate different portions of the beam line when needed.

The last component of the beam line is the detector, which should ideally be a large-area highly-pixelated position-sensitive detector. These requirements are essential for efficient detection capability and, in particular, for studies on single crystals, where data need to be simultaneously collected over a wide portion of the reciprocal space, with their full vectorial $\textbf{Q}$ dependence. The detector should cover also the low angle region, thus allowing studies in the first Brillouin zones, around the (000) reciprocal lattice point. This option is very useful for magnetism in order to study the intensity of the excitations as a function of the total momentum transfer $\textbf{Q}$. This facilitates the distinction of magnetic from nuclear contributions without resorting to polarization analysis. On the other hand, it provides an extremely efficient tool in the case of powder or disordered samples, where averages over the Debye-Scherrer cones can be exploited to greatly enhance the statistical accuracy of the data. To ensure the maximum coverage of the dynamical range limiting also cross-talking effects, the detector should ideally extend up to 1/4 of a sphere. Considering also physical limitations due to the transmitted beam and the shielding, this means an angular coverage from -7.5$^{\circ}$ to 135$^{\circ}$ in the horizontal plane (see Fig. \ref{fig2}) and from -20$^{\circ}$ to 70$^{\circ}$ in the out of plane direction. A good compromise for the sample-to-detector distance is $L_{SD}=4$~m. A shorter distance would indeed result in a broader energy resolution, whereas a longer distance would imply a too large detector area. For $L_{SD}=4$~m and with the above angular coverage, a detector of almost spherical shape results in a surface of about 35~m$^2$, ideally with a typical pixel size of 1~cm$^2$. Much larger pixel sizes might cause a considerable degradation of the $Q$-resolution in the case of small samples.

\section{Instrument performance}
	\label{Section:Performance}
We consider a time-focusing instrument realized as described in Fig. \ref{fig2} and based on a DRCM with HOPG crystals. We can thus estimate its potential performances in terms of wavelength and energy resolution, by comparing the useful flux with that expected for an equivalent chopper spectrometer. Of course, an absolute evaluation of the flux at the sample would require a detailed simulation of the neutron source and the primary neutron transport system, which is far beyond the purpose of this work.   

\subsection{Incident wavelength and resolution}
Referring to Fig. \ref{fig2}, we consider a sample-to-detector distance $L_{SD}=4$~m, a distance between the sample and the exit of the monochromator chamber of 1.5~m, and a 1.5~m long monochromator casemate. This fixes the time-focusing distance $L_{MD}$ between 5.5~m and 7.5~m. As shown in Eq. \ref{Eq:TF4}, once the wavelength is chosen by fixing the Bragg angle and the reflection order, we can match this constraint by varying the rotation frequency. In a steady-state source this frequency can be varied continuously, whereas in a pulsed source it has to be phased to the source frequency, thus introducing a further constraint. However, the combination of Bragg angle, reflection order and rotation frequency provides a really versatile system that can easily span over a wide range of wavelengths. For HOPG crystals, in particular, the peak reflectivity is almost order independent \cite{FreundNIMA1985}, so that even quite high-order reflections can be safely considered.

\begin{figure}[htbp]
	\centering
		\includegraphics[width=0.48\textwidth]{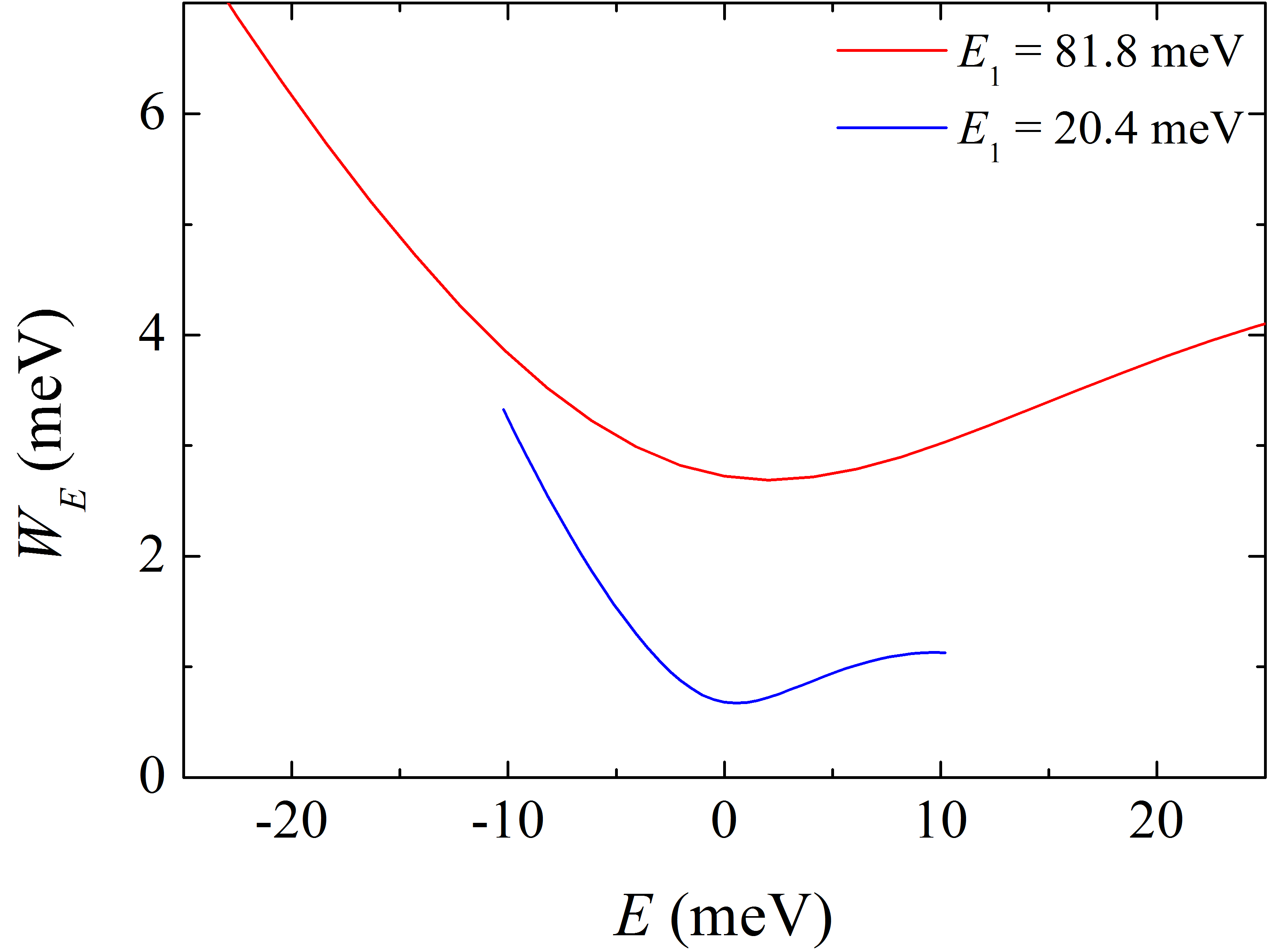}	
		\caption{Instrument energy resolution $W_E$ for two typical configurations. The red line is $W_E$ for 81.8~meV with a monochromator rotation frequency of 84 Hz, using the 4th order of the HOPG crystals with $\theta_M=36.6^{\circ}$. The blue line is obtained for 20.4~meV with a monochromator rotation frequency of 42~Hz at the 2nd order with $\theta_M=36.6^{\circ}$. Both the configurations are detailed in Tab. \ref{Tab:perf1}.}
	\label{fig3}
\end{figure}

Table \ref{Tab:perf1} shows some possible combinations of parameters that match the constraints of the proposed layout for a steady-state source. The table clearly shows the versatility of the instrument that can easily provide incident wavelengths from 1 to 4~\AA~with variable resolution. For long pulse spallation sources like ESS, the monochromator frequency has to be coupled to the source frequency and is then limited to a multiple of the source frequency $\nu_s$. Actually, since both crystal faces create the monochromatic pulse, the rotation frequency can be chosen among the values $n\nu_s/2$, where $n=1,2,3,...$. Fig. \ref{fig3} shows the energy resolution $E$ for two typical instrument configurations.

\begin{table}[htbp]
	\begin{ruledtabular}
		\centering
		\begin{tabular}{cc|ccc|cc}
			$\lambda_i$	&$E_i$ 	&Refl. order		&$\theta_M$			&$\nu_M$		&$L_{MD}$	&$W_E$		\\
			(\AA)				&(meV)	&								&(deg)					&(Hz)				&(m)			&(meV)		\\
			\hline
			1.00				&81.8		&4							&36.6						&84					&5.57			&2.65			\\
			1.00				&81.8		&3							&26.6						&56					&5.62			&3.93  		\\
			\hline
			1.58				&32.7		&3							&45.0 					&70					&5.69			&0.84			\\
			1.58				&32.7		&3							&45.0 					&63					&6.32			&0.93			\\
			\hline
			2.00				&20.4		&3							&63.5						&91					&6.93			&0.33  		\\
			2.00				&20.4		&2							&36.6						&42					&5.57			&0.68			\\
			\hline
			4.00				&5.11		&1							&36.6						&21					&5.57			&0.17			\\
		\end{tabular}
	\caption{Typical instrument configurations based on the design shown in Fig. \ref{fig2}. The elastic energy resolution $W_E$ is obtained using the formulas of Sec. \ref{Section:Resolution}. The table also shows how the same wavelength can be obtained with different combinations of reflection orders and rotation frequencies, e.g. $\lambda_i=1.00$~\AA~and $\lambda_i=2.00$~\AA, or just with different rotation frequencies, e.g. $\lambda_i=1.58$~\AA. Of course this also affects the resolution $W_E$.}
	\label{Tab:perf1}
	\end{ruledtabular}
\end{table}

\subsection{Flux at the sample}
	\label{Section:Flux}
The neutron intensity at the sample can be written in terms of the source flux per unit solid angle and unit energy $\Phi(E_i)$, the guide transmission $T_G$, the incoming beam bi-dimensional divergence accepted by the DRCM $\Delta\Omega_{M}$, the monochromator surface $S_M$, the crystal reflectivity $R$, the neutron pulse frequency $\nu_M$, the DRCM collection time $T$, and the energy window $\Delta E_M$. The intensity $I_S^{TF}$ turns out to be:
\begin{eqnarray}
I_S^{TF} = \Phi(E_i) T_G S_M R^2 \nu_M T \Delta \Omega_{M} \Delta E_M,
	\label{Eq:IntTF}
\end{eqnarray}
\noindent
where the $R^2$ factor is due to the double reflection by the monochromator device. In the real system, the time distribution of neutrons downstream the DRCM is triangular and $T$ is the FWHM.

It is interesting to compare the above formula with the corresponding intensity at the sample for a traditional chopper spectrometer, installed on the same source, that is:
\begin{eqnarray}
I_S^{C} = \Phi(E_i) T_G S_C T_C \nu_C \Delta t \Delta \Omega_{C} \Delta E_C
	\label{Eq:IntC}
\end{eqnarray}
where $S_C$ is the surface of the chopper window opening, $T_C$ is the chopper transmission, $\nu_C$ the neutron pulse frequency, $\Delta t$ the pulse length (FWHM), $\Delta\Omega_{C}$ the bi-dimensional divergence at the chopper position, and $\Delta E_C$ the energy window.

To compare the performance of the proposed time-focusing instrument with a chopper spectrometer we assume the same incident flux $\Phi(E_i)$, ideal transmissions $T_G=T_C=1$ and equal energy windows $\Delta E_M=\Delta E_C$. Moreover, we also have to assume that $\nu_C=\nu_M$, i.e. the two instruments have the same frame overlap and $\Delta t= \eta / (\sqrt{2}\omega_M)$ to ensure the same resolution for the elastic channel. It is worth noting that for pulsed sources this is straightforward since $\nu_C$ and $\nu_M$ are dominated by the source frequency. The gain factor for the time-focusing instrument $G_{TF}$ can be thus written as:
\begin{eqnarray}
G_{TF} = \frac{S_M \Delta \Omega_{M}}{S_C \Delta \Omega_{C}} R^2 \frac{T}{\Delta t}.
	\label{Eq:IntGain}
\end{eqnarray}

The gain factor $G_{TF}$ is thus defined by two critical ratios, namely $(S_M \Delta\Omega_M)/(S_C \Delta\Omega_C)$ and $T/\Delta t$. The first ratio is simply the change in phase-space acceptance of the two instruments, whereas the last term is specifically connected to the time focusing approach and derives from the different manipulation of the wavelength-time volume compared with a chopper instrument.

As already described in Sec. \ref{Section:Instrument} and in Ref. \cite{ ZanattaJPCS2016}, DRCM with a rather large surface can be easily produced -- a cross-sectional area of about 50 cm$^2$ seems reasonable, while the accepted divergence corresponding to the crystal mosaic spread is about $2 \times 10^{-4}$ steradians, assuming that the upstream divergence is at least that large. On the other hand, the construction of a large beam chopper rotating in excess of 100 Hz presents more mechanical constraints. Beam areas at the position of the final, wavelength-resolution-defining chopper are typically of the order of 10 cm$^2$, while the beam divergence for a modern supermirror guide of that guide cross-section is of the order of $2-4 \times 10^{-4}$ steradians, depending on the wavelength. A chopper spectrometer would typically employ a focusing guide after the last chopper, resulting in an increased divergence and a smaller beam spot at the sample, and, at best, a conservation of the phase space at the last chopper position. Consequently, the accepted phase space volume is typically at least twice as large for the case of the DRCM setup, compared to the chopper instrument. 

The term $R^2$ adds a further 0.49 in favor of the chopper instrument, with the result that these parameters globally more or less cancel out, so that the actual discriminating factor is the time-focusing ratio $T/\Delta t$ and in particular the collection time $T$, which is more difficult to evaluate. In principle, at pulsed sources $T$ is limited by the pulse length only. In real cases, this is not completely true and the maximum collection time is defined by the phasing between the first and second monochromators. However, when the rotation angle from the central position is small, the phasing between the two monochromators is preserved and the wavelength change is compensated by the flight path. This compensation is perfect only at first order of the shift angle $\alpha = \pm \omega_M T$. Within this limit, $T$ can be written as:
\begin{eqnarray}
T =\frac{w\sin(\theta_M)\sin(2 \theta_M)}{D \omega_M},
	\label{Eq:IntGain}
\end{eqnarray}
\noindent where $\theta_B$ is the Bragg angle, $D$ the separation between the primary and secondary beam, and $w$ is the length of the monochromator surface illuminated by the beam. In order to estimate the ratio $T/\Delta t$ we consider a monochromatic beam of $\lambda_i=1.01$~\AA, extracted using the (006) reflection of a HOPG-based DRCM. As reported in Tab. \ref{Tab:perf1}, the rotation frequency for the DRCM is $\nu_M=55$~Hz and the Bragg angle $\theta_M=26.6^{\circ}$. Assuming a primary beam of $8\times8$~cm$^2$, we have a monochromator length $w=17.8$~cm and a beam separation $D=1$~m. In this configuration we have $\Delta t=18~\mu$s and $T=185~\mu$s, hence the time-focusing ratio $T/\Delta t$ is about 10. This result was also confirmed by several Monte Carlo simulations of a simple DRCM system.

Furthermore, the DRCM physically separates the primary and the monochromatic beams so that the background is strongly reduced and the signal-to-noise ratio is enhanced. A similar result for a chopper solution is not possible and, to avoid a direct view of the source, a long and curved primary guide is needed.

\section{Conclusions}
	\label{Section:Conclusion}
We have presented a modified version of the time-focusing technique based on a double rotating-crystal monochromator which allows to disentangle time-focusing from beam divergence. This approach is expected to be particularly effective for long-pulse and steady-state neutron sources and, using a longer extraction time and a wider beam cross-section, it can provide a higher flux at the sample with respect to standard chopper instruments, with a gain factor $G_{TF}\sim10$. Moreover, the monochromatic beam is shifted from the primary one, so that the source is always out of line-of-sight from the sample. This ensures a reduced environmental background and enhances the signal-to-noise ratio. Although a spectrometer based on a DRCM can be a rather short instrument, this does not affect the versatility of the instrument in terms of possible incident wavelength and resolutions. The instrument versatility can be further improved by implementing a multiplexing option by adding more DRCMs in the casemate. This is equivalent to the repetition rate multiplication\cite{MezeiPB2000,RussinaNIMA2009} (RRM) that is efficiently employed by chopper spectrometers. The extraction time $T$ can be adjusted by means of appropriate collimators inserted between the two monochromators, thus introducing further flexibility for specific applications. 

If the scientific application can benefit from increased flux over a small sample area, at the expense of an increase in divergence, it should be possible to arrange the crystals in a converging geometry or to phase their rotation in order to obtain spatial focusing effects. Alternatively, or in combination, a focusing guide could be employed between the DRCM setup and the sample. 

However, from the scientific point of view, the combination of low divergence and high flux can be exploited to extend the $(Q,E)$ range of standard spectrometers down to the low-$Q$, low-$E$ region. This region is presently accessible only by specific instruments such as the Brillouin neutron spectrometer BRISP (ILL, Grenoble, France)\cite{AisaNIMA2005,ZanattaRSI2017} and HRC (J-PARC, Japan)\cite{ItohJPSJ2013}, and it has great interest for disordered and magnetic systems. In particular, such an instrument would allow a complete overview of the atomic dynamics, simultaneously probing the whole pattern of low-$Q$ collective excitations and the high-$Q$ response, which provides additional information like the density of states of vibrational modes. This has been proven pivotal for spin waves\cite{IbukaPRB2017}, liquid metals\cite{ZanattaPRL2015}, glasses\cite{ZanattaJPCL2013}, biological systems like DNA\cite{EijckPRL2011} and its hydration water\cite{CornicchiJCP2011}. In addition, for magnetism, INS at low-$Q$ is the only reliable technique to study magnetic excitations in disordered or polycrystalline samples with small magnetic cross-sections, e.g. Ref. \cite{ItohNatComm2016}. As a matter of fact, for $Q\rightarrow0$ the magnetic form factor has its highest value and the vibrational component is minimized \cite{Lovesey}. Conversely, on increasing $Q$ the magnetic contribution disappears and the vibrational one has its maximum. This situation allows for a safe separation of the two components in a single measurement. 
  
\begin{acknowledgments}
The presented results were part of the Tempus Fugit project, which has been supported by Elettra-Sincrotrone Trieste S.C.p.A. through a special grant of the Italian Ministry of Instruction, University and Research (MIUR). FS acknowledges very useful discussions with Jens-Boie Suck and Bruno Dorner on the characteristics of the rotating monochromators used at ILL.
\end{acknowledgments}


\begin{thebibliography}{30}
\bibitem{VettierNIMA2009} C. Vettier, C. J. Carlile, P. Carlsson, Nucl. Instr. and Meth. A \textbf{600}, 8 (2009).
\bibitem{LindroosNIMB2011} M. Lindroos, S. Bousson, R. Calaga, H. Danared, G. Devanz, R. Duperrier, J. Eguia, M. Eshraqi, S. Gammino, H. Hahn, A. Jansson, C. Oyon, S. Pape-M\o ller, S. Peggs, A. Ponton, K. Rathsman, R. Ruber, T. Satogata, G. Trahern, Nucl. Instr. and Meth. B \textbf{269}, 3258 (2011).
\bibitem{OllivierSFN2010} J. Ollivier, J.M. Zanotti, Collection SFN \textbf{10}, 379 (2010)
\bibitem{BewleyNIMA2002} R. Bewley, R. Eccleston, Nucl. Instr. Meth. Phys. Res. A \textbf{492}, 97 (2002).
\bibitem{EhlersRSI2011} G. Ehlers, A. Podlesnyak, J.L. Niedziela, E.B. Iverson, Rev. Sci. Instrum. \textbf{82}, 085108 (2011).
\bibitem{StoneRSI2014} M.B. Stone, J.L. Niedziela, D.L. Abernathy, L. de-Beer Schmitt, G. Ehlers, O. Garlea, G.E. Granroth, M. Graves-Brookes, A.I. Kolesnikov, A. Podlesnyak and B. Winn, Rev. Sci. Instrum. \textbf{85}, 045113 (2014).
\bibitem{KajimotoJPSJ2010} R. Kajimoto,  M. Nakamura,  Y. Inamura,  F. Mizuno,  S.-K. K.Nakajima,  T. Yokoo, T. Nakatani, R. Maruyama, K. Soyama, K. Shibata, K.Suzuya, S. Sato, K. Aizawa, M. Arai, S. Wakimoto, M. Ishikado, S. Shamoto, M. Fujita, H. Hiraka, K. Ohoyama, K. Yamada, and C.-H. Lee, J. Phys. Soc. Jpn. \textbf{80}, SB025 (2010).
\bibitem{NakajimaJPSJ2011} K. Nakajima, S. Ohira-Kawamura, T. Kikuchi, M.Nakamura, R. Kajimoto, Y. Inamura, N. T. andd K. Aizawa, K. Suzuya, K.Shibata, T. Nakatani, K. Soyama, R. Maruyama, H. Tanaka, W. Kambara, T. Iwahashi, Y. Itoh, T. Osakabe, S.Wakimoto, K. Kakurai, F. Maekawa, M. Harada, K. Oikawa, R. E. Lechner, F.Mezei, and M. Arai, J. Phys. Soc. Jpn. \textbf{80}, SB028 (2011).
\bibitem{ItohJPSJ2013} S. Itoh, T. Yokoo, D. Kawana, Y. Endoh, J. Phys. Soc. Japan \textbf{82}, SA034 (2013).
\bibitem{VoigtNIMA2014} J. Voigt, N. Violini and T. Br\"uckel, Nucl. Instr. Meth. Phys. Res. A \textbf{791}, 26 (2014).
\bibitem{DeenJPCS2015}  P.P. Deen, A. Vickery, K.H. Andersen and R. Hall-Wilton, J. Phys. Conf. Ser. \textbf{83}, 03002 (2015).
\bibitem{CicognaniPB2000} G. Cicognani, H. Mutka, D. Weddle, B. Hamelin, P. Malbert, F. Sacchetti, C. Petrillo, E. Babucci, Physica B \textbf{276}, 85 (2000).
\bibitem{Blanc1983} Y. Blanc, ILL Int. Rep. 83BL21G (1983).
\bibitem{JuranyiCP2003} F. Jurányi, S. Janssen, J. Mesot, L. Holitzner, C. K\"agi, R. Tuth, R. Bürge, M. Christensen, D. Wilmer, R. Hempelmann, Chem. Phys. \textbf{292}, 495–499 (2003).
\bibitem{ZanattaJPCS2016} M. Zanatta, A. Orecchini, S. Aisa, F. Casinini, L. Farnesini, P.P. Deen, A. Paciaroni, C. Petrillo, F. Sacchetti, J. Phys. Conf. Ser. \textbf{746}, 012002 (2016).
\bibitem{FreundNIMA1985} A.K. Freund,  Nucl. Instr. and Meth. A \textbf{238}, 570 (1985) .
\bibitem{WillisAC1960} B.T.M. Willis, Acta Cryst. \textbf{13}, 763 (1960). 
\bibitem{HohlweinJAC1988} D. Hohlwein, D. P. Siddons and J. B. Hastings, J. Appl. Cryst. \textbf{21} 911-915 (1988).
\bibitem{RisteNIMA1969} T. Riste, K. Otnes, Nucl. Instr. and Meth. A \textbf{75}, 197 (1969).
\bibitem{VioliniNIMA2014} N. Violini, J. Voigt, S. Pasini, T. Br\"uckel, Nucl. Instr. and Meth. A \textbf{736}, 31–39 (2014) and references therein.
\bibitem{MezeiPB2000} F. Mezei, M. Russina, S. Schorr, Physica B \textbf{276}, 128 (2000)
\bibitem{RussinaNIMA2009} M. Russina, F. Mezei, Nucl. Instrum. and Meth. A \textbf{604}, 624 (2009).
\bibitem{AisaNIMA2005} D. Aisa, E. Babucci, F. Barocchi, A. Cunsolo, F. D'Anca, A. De Francesco, F. Formisano, T. Gahl E. Guarini, S. Jahn, A. Laloni, H. Mutka, A. Orecchini, C. Petrillo, F. Sacchetti, J.B. Suck and G. Venturi, Nucl. Instr. Meth. Phys. Res. A \textbf{544}, 620-642 (2005).
\bibitem{ZanattaRSI2017} M. Zanatta, F. Barocchi, A. De Francesco, E. Farhi, F. Formisano, E. Guarini, A. Laloni, A. Orecchini, A. Paciaroni, C. Petrillo, W-C. Pilgrim, J-B. Suck, F. Sacchetti, Rev. Sci. Instrum. \textbf{8}, 053905 (2017) and references therein.
\bibitem{IbukaPRB2017} S. Ibuka, S. Itoh, T. Yokoo, and Y. Endoh, Phys. Rev. B \textbf{95}, 224406 (2017).
\bibitem{ZanattaPRL2015} M. Zanatta, F. Sacchetti, E. Guarini, A. Orecchini, A. Paciaroni, L. Sani, C. Petrillo, Phys. Rev. Lett. \textbf{114}, 187801 (2015).
\bibitem{ZanattaJPCL2013} M. Zanatta, A. Fontana, A. Orecchini, C. Petrillo, and F. Sacchetti, J. Phys. Chem. Lett. \textbf{4}, 1143-1147 (2013).
\bibitem{EijckPRL2011} L. van Eijck, F. Merzel, S. Rols, J. Ollivier, V.T. Forsyth, and M.R. Johnson, Phys. Rev. Lett. \textbf{107}, 088102 (2011).
\bibitem{CornicchiJCP2011} E. Cornicchi, F. Sebastiani, A. De Francesco, A. Orecchini, A. Paciaroni, C. Petrillo, F. Sacchetti, J. Chem. Phys. \textbf{135}, 	025101 (2011).
\bibitem{ItohNatComm2016} S. Itoh, Y. Endoh, T. Yokoo, S. Ibuka, J. Park, Y. Kaneko, K.S. Takahashi, Y. Tokura, and N. Nagaosa, Nat. Commun. \textbf{7}, 11788 (2016).
\bibitem{Lovesey} S.W. Lovesey. Theory of Neutron Scattering from Condensed Matter. Volume II. (Oxford University Press, 1986).
\end{thebibliography}
\end{document}